\documentclass[aps,prb,twocolumn,showpacs,superscriptaddress,floatfix]{revtex4-1} 
\usepackage{subfigure,amsmath,graphicx,epsfig,amssymb}
\usepackage{dcolumn}   
\usepackage{bm}  
\usepackage{epstopdf}       
\usepackage{amssymb}
\usepackage{amsmath}   
\usepackage{chngcntr}

\begin{document}

\title{Correlating the Energetics and Atomic Motions of the Metal-Insulator Transition of M$_{1}$ Vanadium Dioxide}
\author{J.~M.~Booth}
\email{jamie.booth@rmit.edu.au}
\affiliation{Theoretical Chemical and Quantum Physics, School Science, RMIT University, Melbourne VIC 3001, Australia}

\author{D.~W.~Drumm}
\affiliation{Theoretical Chemical and Quantum Physics, School Science, RMIT University, Melbourne VIC 3001, Australia}

\affiliation{Australian Research Council Centre of Excellence for Nanoscale BioPhotonics, Applied Physics, School of Applied Sciences, RMIT University, Melbourne 3001, VIC, Australia}

\author{P.~S.~Casey}
\affiliation{CSIRO Manufacturing, Clayton VIC 3168, Australia}

\author{J.~S.~Smith}
\affiliation{Theoretical Chemical and Quantum Physics, School Science, RMIT University, Melbourne VIC 3001, Australia}

\author{A. ~J. ~Seeber}
\affiliation{CSIRO Manufacturing, Clayton VIC 3168, Australia}

\author{S. ~K. ~Bhargava}
\affiliation{Centre for Advanced Materials and Industrial Chemistry, School Science, RMIT University, Melbourne VIC 3001, Australia}

\author{S.~P.~Russo}
\affiliation{Theoretical Chemical and Quantum Physics, School Science, RMIT University, Melbourne VIC 3001, Australia}

\date{\today}

\begin{abstract}
Materials that undergo reversible metal-insulator transitions are obvious candidates for new generations of devices. For such potential to be realised, the underlying microscopic mechanisms of such transitions must be fully determined. In this work we probe the correlation between the energy landscape and electronic structure of the metal-insulator transition of vanadium dioxide and the atomic motions occurring using first principles calculations and high resolution X-ray diffraction. Calculations find an energy barrier between the high and low temperature phases corresponding to contraction followed by expansion of the distances between vanadium atoms on neighbouring sub-lattices. X-ray diffraction reveals anisotropic strain broadening in the low temperature structure's crystal planes, however only for those with spacings affected by this compression/expansion.  GW calculations reveal that traversing this barrier destabilises the bonding/anti-bonding splitting of the low temperature phase. This precise atomic description of the origin of the energy barrier separating the two structures will facilitate more precise control over the transition characteristics for new applications and devices.
\end{abstract}

\maketitle
\section*{Introduction}

The reversible phase transition of VO$_{2}$ at $\sim$ 340 K occurs between a low temperature, insulating monoclinic structure, and a high temperature, metallic tetragonal form.\cite{Morin1959, Zylberstein1975,Eyert2002} The transition between between the insulating and metallic forms results in a switch from transparent to absorbing in the near infra-red,\cite{Verleur1968,Zylberstein1975,Eyert2002} which can occur on time-scales as low as femtoseconds when triggered by laser pumping.\cite{Cavalleri2001} While this transition was first identified by Morin in 1959,\cite{Morin1959} and explored more thoroughly in the 1970s by authors such as Goodenough,\cite{Goodenough1971} Pouget\cite{Pouget1974} and Mott,\cite{Zylberstein1975} the last decade has seen an explosion of research into devices based upon this transition.\cite{Wei2009,Nakano2012,Wei2012,liu2012,Park2013} The trigger for this has in part been the maturation of fabrication procedures which allow nanostructures of vanadium dioxide to be grown and utilized.\cite{Cao2009,Wu2013}

Much of the existing theoretical research has been devoted to answering the question of whether the insulating form of VO$_2$ is a band- or a Mott-Hubbard insulator,\cite{Zylberstein1975,Wentzcovich1994,Eyert2002,Tomczak2008} in an effort to determine the roles of both correlations and lattice symmetry breaking in the transition, with most work confirming that both effects are important.\cite{Biermann2005,Tomczak2007,Gatti2007,Tomczak2008,Belozerov2012} However, a complete description of the interplay between the energetics, the atomic rearrangements and the electronic structure of VO$_2$ as it transitions between the monoclinic and tetragonal structures has remained elusive. Device design and optimization requires detailed knowledge of the energy landscape across the transition with respect to changes in the lattice structures. This knowledge has become particularly important in recent years with the development of devices based upon the modulation of the metal-insulator transition of VO$_2$ by inputting stress or strain.\cite{Sohn2009,Cao2009,Jones2010,Aetukuri2013,Park2013} 

Evidence for a soft mode connecting the tetragonal to the monoclinic structures was found as far back as 1978 by Terauchi and Cohen,\cite{Terauchi1978} who found a lattice instability at the $R$ point of the tetragonal structure using diffuse X-ray scattering. Gervais and Kress\cite{Gervais1985} used a shell model to calculate the phonon dispersion curves of the tetragonal form of VO$_{2}$, and also found a softening of the lowest frequency mode at the $R$ point. 

Beginning with the work of Cavalleri et al.\cite{Cavalleri2001} pump-probe measurements have shed considerable light on the lattice dynamics occurring across the transition. A structural ``bottleneck" associated with the phonon connecting the monoclinic and tetragonal structures was observed upon hole photo-doping,\cite{Cavalleri2004} suggesting that the insulating phase depends significantly on the lattice potential, indicating band-like character. Kim et al.\cite{Kim2006} used pump-probe measurements in conjunction with X-ray diffraction and found that the sharp resonance corresponding to the monoclinic A$_{g}$ peak disappears at the transition with lower energy and less intense tetragonal B$_{g}$ resonances replacing them. Wall et al.\cite{Wall2012} also used pumping to modify the lattice local potential and examine its effects on the coherent phonon spectrum of VO$_{2}$, as an example of the general applicability of the use of pumping to induce a change in lattice potential, which can be used to study relaxation processes. However, a theoretical description of the interplay between the lattice potential and the atomic and electronic structure has proven elusive.
\begin{figure*}[th!]
  \includegraphics[width=1.7\columnwidth]{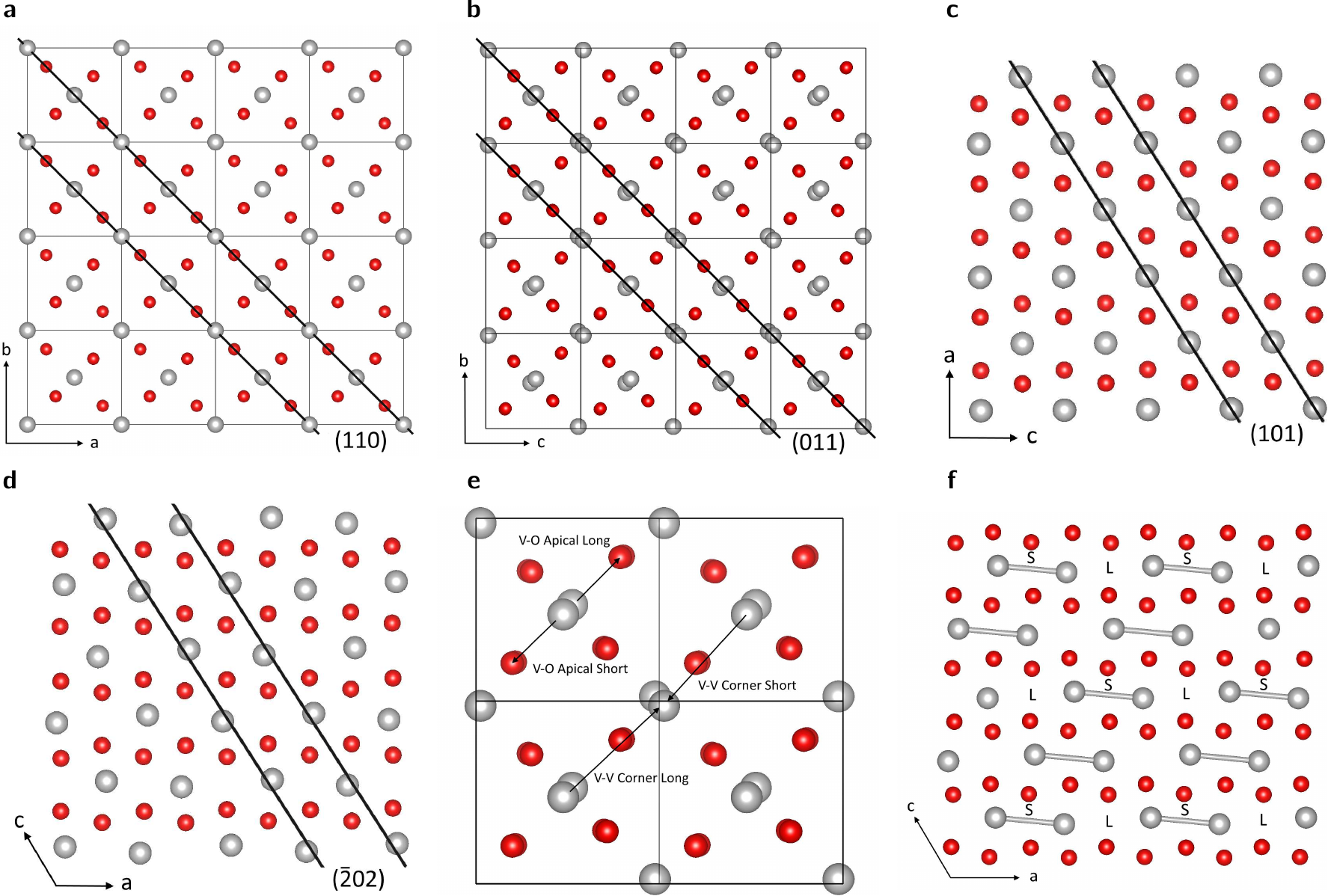}
  \caption{a) view down $\langle$001$\rangle$ of the tetragonal structure, the (110) planes are indicated by black lines, b) view down $\langle$100$\rangle$ M$_{1}$ structure, the (011) planes are indicated by black lines, illustrating that they correspond to the same distance as the tetragonal (110) planes, the off-centre positioning of the vanadium atoms in the M$_{1}$ structure from the anti-ferroelectric twist is also apparent c) view down $\langle$010$\rangle$ of the tetragonal structure, the evenly spaced vanadium chains are visible, running parallel to the c-axis, and the (101) planes are indicated with black lines, d) view down $\langle$010$\rangle$ of the M$_{1}$ structure, with the Peierls paired vanadium chains visible running parallel to the a-axis and the ($\bar{2}02$) planes are marked with black lines. These are shifted by one half of a lattice spacing for better comparison to the tetragonal (101) planes, illustrating that both correspond to the same distance, e) same perspective as b) but with the ``V-V Corner Long", ``V-V Corner Short", ``V-O Apical Long" and ``V-O Apical Short" distances marked and f) same perspective as d) but with the ``V-V Chain Short" and ``V-V Chain Long" distances indicated by the letters ``S" and ``L" respectively.} 
\end{figure*}

The computational study of Zheng and Wagner\cite{Zheng2015} utilised the Quantum Monte Carlo approach to show that the MIT is a direct consequence of the change in structure, that is the monoclinic structure is insulating, and the tetragonal form exhibits metallic behaviour. While sounding trivial, there has been some conjecture over the coincidence of the structural and electronic phase transitions,\cite{Laad2006,Kim2006} which Zheng and Wagner, and also this work resolve. Chen et al.\cite{Chen2015} explored the properties of the parameter space spanned by the $\beta$ angle and the tetragonal c-axis using the DFT+U approach,\cite{Anisimov1991} and suggested that changing orbital occupancy is initially responsible for opening the band gap as a result of dimerisation, which is widened by a subsequent increase in the antiferroelectric distortion.

Thus what is missing from the literature as it currently stands is an exploration of the energy landscape of the structural phase transition with respect to the metal-insulator transition in terms of exactly what constitutes the separation between the two structures. The intent of this work is to utilize a comprehensive computational approach to determining the processes occurring during the metal-insulator and structural phase transitions, and to combine it with experimental data to confirm the predictions of our calculations. Specifically, the outstanding questions we seek to answer are: i) literature data suggests a latent heat of $\sim$ 40 J/g for the transition,\cite{Booth2009} to what does this energy barrier correspond? Which particular atomic motions give rise to this barrier, ii) if a minimum energy path can be mapped between the structures, and the aforementioned atomic displacements determined, what are the effects of these displacements on the electronic structure? Are the structural phase transitions and metal-insulator transitions necessarily coincident as suggested by Zheng and Wagner?\cite{Zheng2015} 

We start by computing the lowest energy path between the structures using the nudged elastic band technique\cite{Henkelman2000} and density functional theory to determine this energy landscape. The DFT data reveal that in order for the structural transition to occur, the inter-vanadium spacing along the [110] or [1$\bar{1}$0] directions must be compressed, generating electronic repulsion and thus an energy barrier. High resolution X-ray diffraction measurements reveal anisotropic strain related to the atomic spacing in these directions in the monoclinic structure, which is not present in the tetragonal form. Frequency-dependent GW calculations reveal that the top of the barrier corresponds to the opening of the gap due to bonding/anti-bonding splitting as the vanadium atoms dimerise. The data indicate that the most efficient modulations of the transition temperature involve stress input along [110] or [1$\bar{1}$0] of the tetragonal structure or the [011] or [01$\bar{1}$] directions of the monoclinic structure, consistent with the action of doping with tungsten.\cite{Booth2009a}

\begin{figure}[h!]
  \includegraphics[width=0.8\columnwidth]{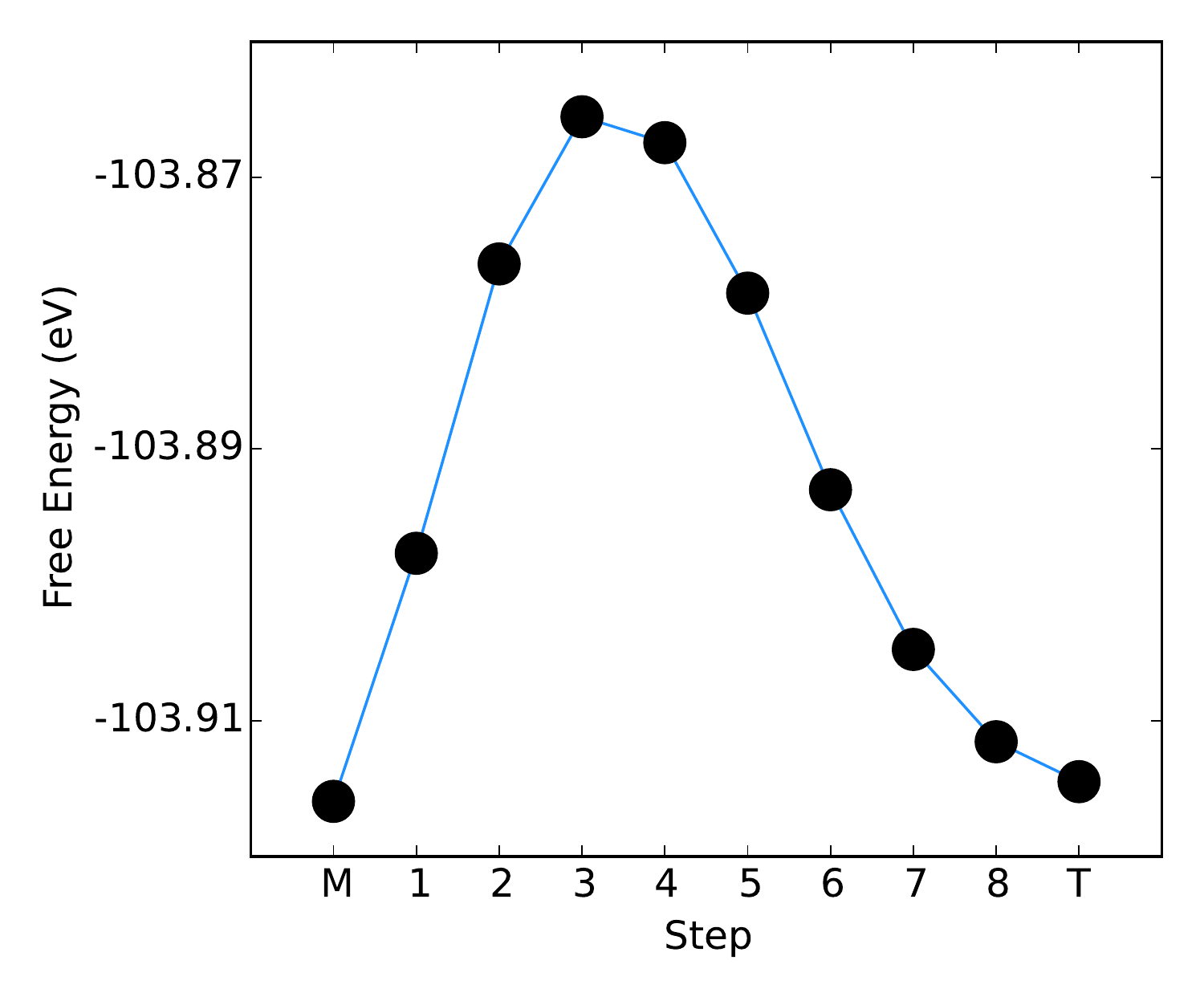}
  \caption{Total free energies (eV, black filled circles) of the structures across the transition from a combination of DFT geometry relaxations and the elastic band method. M and T correspond to the monoclinic and tetragonal structures respectively, while the intermediate structures are denoted by step numbers.}     
\end{figure}
\begin{figure*}
  \includegraphics[width=1.7\columnwidth]{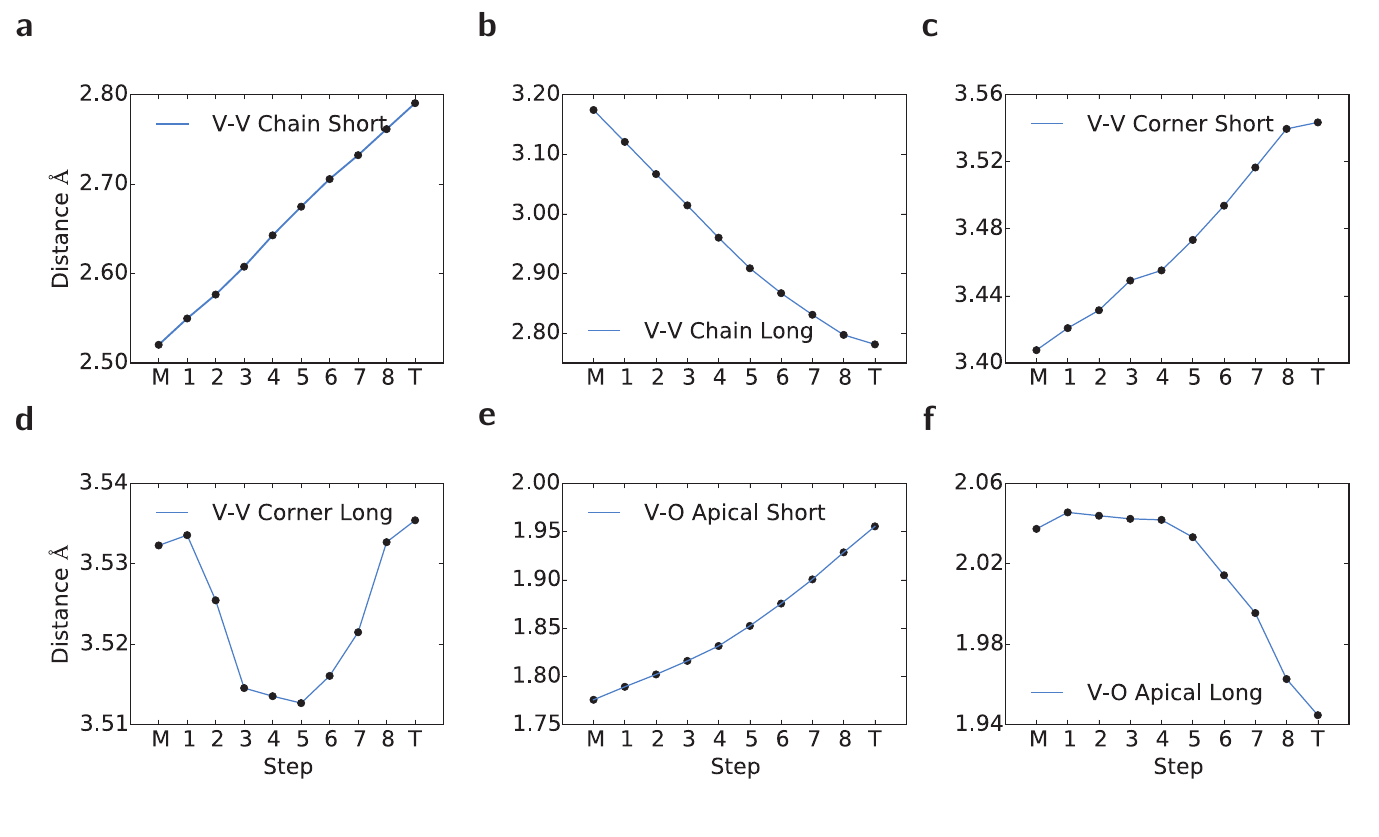}
  \caption{Variation of characteristic distances of the VO$_{2}$ structure across the energy path determined by the elastic band calculations. The data points (filled black circles) are linearly interpolated to guide the eye. Of note are monotonic trends with step progression of all distances apart from V-V Corner long, which initially contracts, then expands. Illustrations of what each distance corresponds to are contained in Figure 1.} 
\end{figure*}
\section*{Results}
\subsection*{Structural Rearrangements}
The relevant structural characteristics of tetragonal and M$_{1}$ VO$_{2}$ are presented in Figure 1. Figures 1a and 1b compare the view of the tetragonal structure down $\langle001\rangle_{T}$ (the subscript T or M refers to tetragonal or monoclinic respectively) with the view of the M$_{1}$ structure down $\langle100\rangle_{M}$. The comparison illustrates that the structural rearrangements occurring in the transition from the tetragonal to the monoclinic structure orthogonal to the monoclinic a-axis can be summarized as an alternating off-set of the vanadium atoms from the centers of the oxygen octahedra. This off-set occurs in the long axis of the octahedra. The (110)$_{T}$ and (011)$_{M}$ planes of the tetragonal and monoclinic structure are also indicated with black lines, which reveals that they correspond to equivalent atomic spacings in each structure (although due to a slight expansion of the tetragonal structure its diffraction peak manifests at slightly lower angle).

Figures 1c-d present a comparison of the tetragonal and M$_{1}$ structures down $\langle010\rangle$ (this axis is coincident for the tetragonal and monoclinic structures), which indicates that the changes occurring across the structural phase transition parallel to the monoclinic a-axis consist of the evenly spaced vanadium atoms (the ``vanadium chains") of the tetragonal structure pairing up (the so-called Peierls pairing), forming an alternating long-short pattern of inter-vanadium spacing. The (101)$_{T}$ and ($\bar{2}$02)$_{M}$ planes are indicated in the tetragonal and monoclinic structures respectively, the ($\bar2$02)$_{M}$ plane has been shifted by one half of its spacing to illustrate that it is the equivalent distance in the monoclinic structure of the (101)$_{T}$ plane.

Figure 1e illustrates the four characteristic distances of interest in this study which are orthogonal to the monoclinic a-axis. The V-O Apical Long V-O Apical Short distances describe the amount to which the vanadium atoms are off-set from the center of the octahedron; if the numbers are equal then the atom sits at the center of the oxygen octahedron. The V-V Corner Long and Short distances describe the two shortest distances between the vanadium atoms on neighboring chains, these distances lie parallel to the (101)$_{T}$ and (200)$_{M}$, ($\bar{2}01$)$_{M}$ planes respectively. Figure 1f defines the V-V Chain Long (indicated by the letter ``L") and short (``S") distances. These are the distances between the vanadium atoms in the chain which undergoes Peierls pairing. If both distances are equal, as in the tetragonal structure, the vanadium atoms are evenly spaced. As the tetragonal structure transforms into the monoclinic form, the atoms pair up and one of these distances decreases, while the other increases.

\begin{figure*}[]
  \includegraphics[width=2.0\columnwidth]{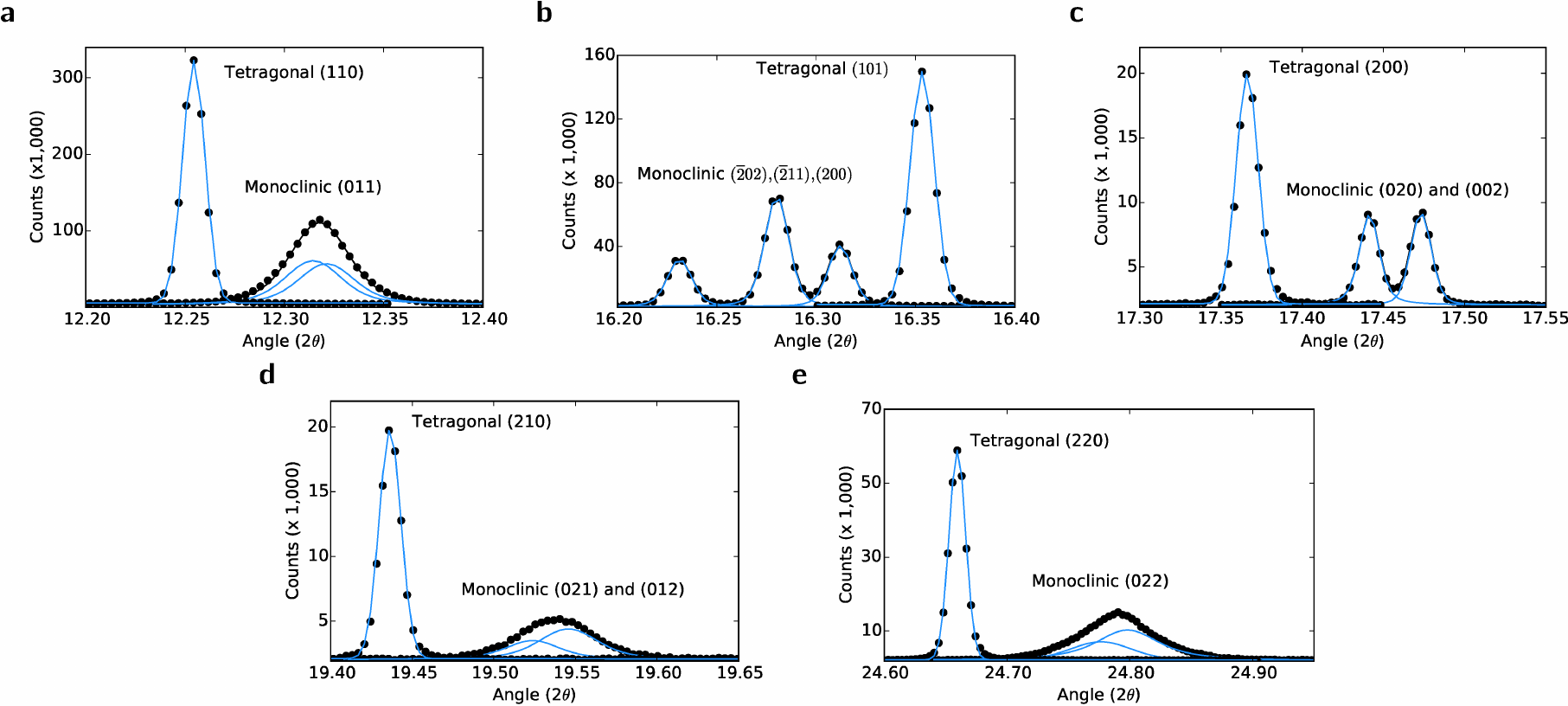}
  \caption{X-ray diffraction data and corresponding fits of a) the monoclinic (011) and tetragonal (110) peaks, b) the monoclinic ($\bar{2}$02), ($\bar{2}$11) and (200) peaks and the corresponding tetragonal  (101) peak, c) the monoclinic (020) and (002) peaks and the corresponding tetragonal (200) peak, d) the monoclinic (012) and (021) peaks and the corresponding tetragonal (210) peak, and e) the monoclinic (022) and corresponding tetragonal (220) peaks.}     
\end{figure*}

\subsection*{Nudged Elastic Band Calculations}
The total free energies of the structures obtained along the minimum energy path between the monoclinic and tetragonal structures determined by the nudged elastic band method are plotted in Figure 2. While the energies of the monoclinic and tetragonal structures are almost degenerate, there is a clear energy barrier between the two structures. This corresponds to an energy of 18.6 J/g, which compares favorably with the experimentally observed specific heat of the phase transition of $\sim$ 40 J/g.\cite{Booth2009} However, this result poses the obvious question: to what does this barrier correspond to in terms of the structural rearrangements?
Figures 3a-f plot the a) V-V Chain Short (aka Peierls) spacing, b) the V-V Chain long distance, c) the V-V Corner Short distance, d) the V-V Corner Long distance, e) the V-O Apical Short and f) V-O Apical Long distance. Figures 3a-b indicate that the Peierls pairing distance increases continuously across the transition from monoclinic to tetragonal, while unsurprisingly, the corresponding long inter-vanadium distance decreases monotonically. This data simply expresses the fact that the evenly spaced vanadium atom chains running along the tetragonal c-axis (monoclinic a-axis) experience a Peierls distortion and adopt a long-short internuclear spacing configuration. The monotonicities of the plots do not suggest an origin for the energy barrier of Figure 2.

Figures 3c-d however tell a different story. Figure 3c indicates that the short distance between the central vanadium atoms and the corner vanadium atoms (see Figure 1c) increases monotonically across the transition, however the longer distance (plotted in Figure 3d) initially contracts, and then expands. Comparison of Figures 2 and 3 suggest that the peak of the total energy corresponds approximately to the minimum in the long V-V corner distance. Figures 3e-f illustrate the trends of the apical vanadium-oxygen distances, and the shorter distance again displays a monotonic increase across the metal-insulator transition, however the longer distance initially plateaus, before decreasing significantly.

Thus the only behavior occurring across the metal-insulator transition consistent with an energy barrier, i.e. an initial increase and subsequent decrease, is the compression and expansion of the long V-V corner distance. This indicates that the force needed to effect the transition between the structures is directed approximately along the diagonals of the unit cell, perpendicular to the vanadium chains. This corresponds to the [110]$_{T}$ and [1$\bar{1}$0]$_{T}$ directions of the tetragonal structure. These directions describe the spacing of the \{110\} planes of the tetragonal structure, and the \{011\} planes of the monoclinic structure. Such an effect closely mirrors the observations of Pouget \textit{et al.}\cite{Pouget1975} who found that inputting a uniaxial stress along the [110]$_{T}$ direction resulting in the appearance of the M$_{2}$ monoclinic form of vanadium dioxide. X-ray absorption also revealed that changes in this distance in tungsten-doped VO$_{2}$ correlated with the amount of tungsten doped into the lattice and therefore the degree to which the transition temperature was depressed.\cite{Booth2009a} Thus this direction seems to be of significance in the structural phase transition. Investigation of any changes in these spacings occurring may therefore confirm the prediction of the computational approach.

\begin{table}[h!]
\centering
	\begin{tabular}{l| c }
	\hline
	\hline
	Plane & $\sigma$ \\
	\hline
	(110) & 5.7$\times$10$^{-3}$ \\
	(101) & 6.1$\times$10$^{-3}$  \\
	(200) & 6.8$\times$10$^{-3}$  \\
	(210) & 6.7$\times$10$^{-3}$  \\
	(220) & 6.7$\times$10$^{-3}$ \\
	\hline
	
	\end{tabular}	
	\caption{Gaussian broadening parameters ($\sigma$) of the fits to the diffraction peaks of the tetragonal structure.}
\end{table}
\begin{table}[h!]
\centering
	\begin{tabular}{l| c| c }
	\hline
	\hline
	Plane & $\sigma$ &  $\Gamma$ \\
	\hline
	(011) & 1.05$\times$10$^{-2}$ & 7.7$\times$10$^{-3}$ \\
	(011) & 1.03$\times$10$^{-2}$ & 7.8$\times$10$^{-3}$ \\
	($\bar{2}$02) & 5.2$\times$10$^{-3}$ & 1.9$\times$10$^{-3}$ \\
	($\bar{2}$11) & 5.2$\times$10$^{-3}$ & 1.8$\times$10$^{-3}$ \\
	(200) & 5.2$\times$10$^{-3}$ & 1.7$\times$10$^{-3}$ \\
	(020) & 4.4$\times$10$^{-3}$ & 3.2$\times$10$^{-3}$ \\
	(002) & 5.7$\times$10$^{-3}$ & 1.5$\times$10$^{-3}$ \\
	(021) & 1.04$\times$10$^{-2}$ & 1.17$\times$10$^{-2}$ \\
	(012) & 1.2$\times$10$^{-2}$ & 1.02$\times$10$^{-2}$ \\
	(022) & 2.08$\times$10$^{-2}$ & 9.7$\times$10$^{-3}$ \\
	(022) & 1.08$\times$10$^{-2}$ & 2.5$\times$10$^{-2}$ \\
	\hline
	
	\end{tabular}	
	\caption{Gaussian ($\sigma$) and Lorentzian ($\Gamma$) broadening parameters of the Voigt fits to the diffraction peaks of the monoclinic structure.}
\end{table}
\begin{figure*}[]
  \includegraphics[width=1.7\columnwidth]{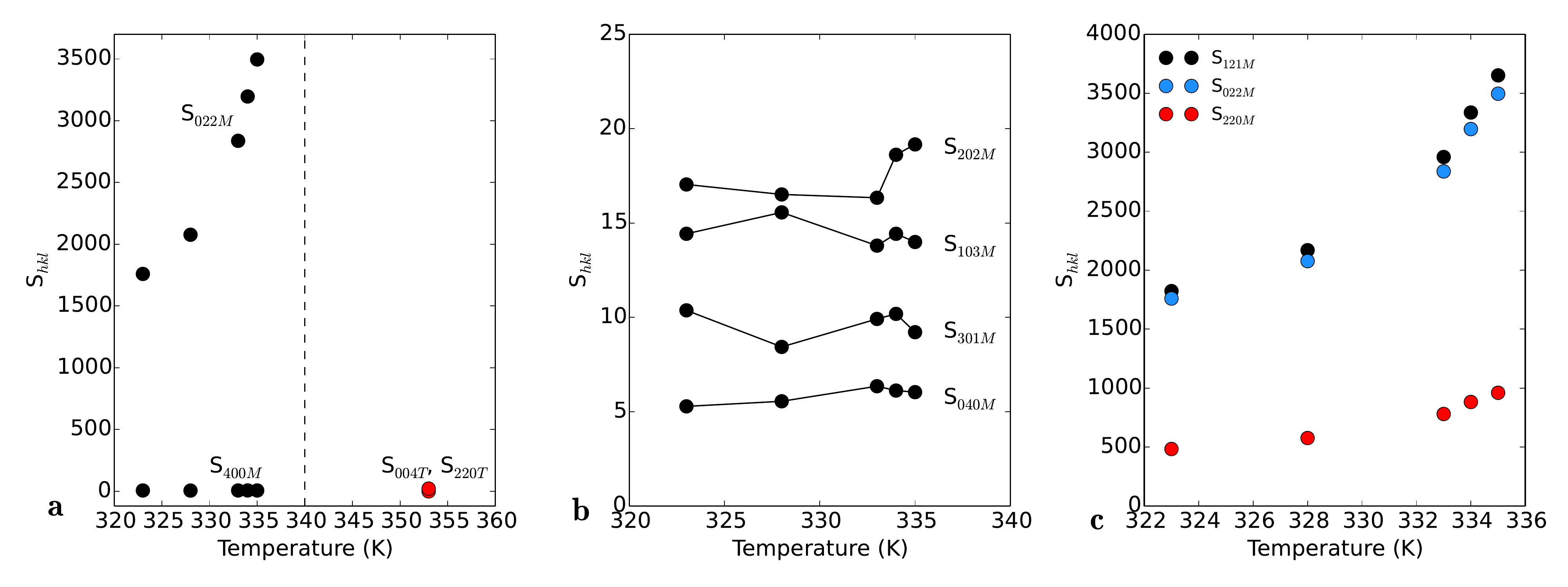}
  \caption{a) Temperature-dependent comparison of strain broadening parameters extracted using the Stephens method. S$_{022M}$ and S$_{400M}$ correspond to the covariances of the $B$ and $C$ reciprocal axes and the variance of the $A$ axis for the Monoclinic structure respectively and are plotted in black. S$_{220T}$ and S$_{400T}$ are the equivalent parameters for the Tetragonal structure, plotted in red. A clear divergence in the covariance of the Monoclinic $B$ and $C$ axes (S$_{022M}$) is observed as the temperature approaches the critical point ($\sim$ 340 K, indicated by a dashed line). b) Comparison of the Monoclinic Stephens strain parameters with no component along the Monoclinic \textbf{b}-axis and c) Comparison of the S$_{121M}$, S$_{022M}$ and S$_{202M}$ parameters, illustrating that while there is some correlation between broadening involcing the \textbf{a}- and \textbf{c} axes, it is far smaller than the broadening involving both the \textbf{b}- and \textbf{c}-axes.}
\end{figure*}
\subsection*{High Resolution X-ray Diffraction}
To determine if there was any manifestation of the effects of this distortion in experimental data, high resolution X-ray diffraction was performed using the powder diffraction beamline at the Australian Synchrotron. Diffraction data of a sample of pure VO$_{2}$ were recorded above and below the structural phase transition temperature of $\sim$ 340 K, and Figure 4 illustrates the most significant properties. Figure 4a contrasts the (110)$_{T}$ and (011)$_{M}$ peaks respectively (which correspond to the same planes, see Figures 1a-b). The data shows clearly that the transition from tetragonal to monoclinic results in slight splitting and considerable broadening of the peaks corresponding to the inter-vanadium distances along the [110]$_{T}$ and [1$\bar{1}$0]$_{T}$ directions.

However, this broadening is not uniform. Figure 4b illustrates the (101)$_{T}$ peak, and a triplet corresponding to the ($\bar{2}$02)$_{M}$, ($\bar{2}$11)$_{M}$, and (200)$_{M}$ peaks (from low to high angle). In this case, the (101)$_{T}$ and ($\bar{2}$02)$_{M}$ peaks correspond to the same distance, and despite a difference in amplitude, the peak shapes are very similar. Therefore, the data indicates that these distances do \textit{not} experience any broadening as the structure transforms from tetragonal to monoclinic.

Figures 4c-e confirm this preferential orientation: Figure 4c contrasts the (200)$_{T}$ and (020)$_{M}$,(002)$_{M}$ peaks (again, which describe the same spacing) and similarly to Figure 4b, no broadening is apparent. Figure 4d however, which contrasts the (210)$_{T}$ and the (021)$_{M}$,(012)$_{M}$ peaks \textit{does} exhibit the broadening observed in the (110)$_{T}$ to (011)$_{M}$ transition of Figure 4a. Figure 4e provides confirmation of the data of Figure 4a: it corresponds to the peaks at half the spacing: (220)$_{T}$ and (022)$_{M}$, and as before, broadening is observed.

Tables I and II presents the fit parameters of the tetragonal and monoclinic peaks of Figure 4 respectively, which confirms this trend; the tetragonal peaks exhibit Gaussian broadening, however it is almost constant across the spectrum, varying only by a maximum of 10\% from the mean. The monoclinic data on the other hand shows systematic variation. While the ($\bar{2}$02)$_{M}$, ($\bar{2}$11)$_{M}$, (200)$_{M}$, (020)$_{M}$ and (002)$_{M}$ peaks exhibit roughly similar Gaussian broadening to the tetragonal peaks and a small amount of Lorentz broadening, the (011)$_{M}$, (021)$_{M}$, (012)$_{M}$ and (022)$_{M}$ are far broader. They exhibit Gaussian broadening which is approximately twice that of the tetragonal and other monoclinic peaks, and Lorentz broadening which is in some cases an order of magnitude larger, for example the (200) and (021) peaks. 

Thus the data indicates that peaks of the form (0$xx$) or (0$xy$) experience significantly more broadening than other orientations. Such spacings describe distances with the same orientation as that of Figure 3d; directed toward the neighboring vanadium chain. This disorder may be reconciled with the NEB data by taking into account the effects of defects and grain boundaries in the structure of the experimental sample. Figure 3d indicates that a distance is initially compressed, and subsequently extends. Figure 2 suggests that this compression costs energy. Thus, if structural defects are present which allow dissipation of this energy, the transformation to the monoclinic structure may be incomplete. Obviously individual grain boundaries will place limits on the extent of the propagation of this, and therefore it is possible that due to this, the strain broadening of the individual grains comprise a distribution which is anisotropic in nature.
To investigate the possible manifestation of this, anisotropic strain broadening parameters were extracted using the Stephens method.\cite{Stephens1999} Figure 5a plots the magnitudes of the S$_{022}$ and S$_{040}$ contributions to the broadening for the M$_{1}$ structure for five temperatures below the critical point, and contrasts them with the equivalent Tetragonal S$_{400}$ and S$_{220}$ contributions respectively (the mismatch in indices between the monoclinic and tetragonal parameters is due to the aforementioned different naming conventions of crystallographic axes in the cells, thus S$_{040M}$ = S$_{400T}$ and S$_{022M}$ = S$_{220T}$). The co-existence of the monoclinic and tetragonal structures near the critical temperature creates issues for fitting, and thus the data of Figure 5 is limited to those points near the transition temperature which exhibited the best fitting parameters.

While the S$_{040M}$ data is independent of temperature, and approximately equivalent in magnitude to the S$_{400T}$ data, the S$_{022M}$ data diverges as the temperature approaches the critical point. This contrasts sharply with the S$_{220T}$ data which is almost \textit{un-}correlated, as S$_{220T}$ = -1.4, which is very close to zero. This data therefore indicates that as the critical temperature is approached from below, the contributions to peak broadening from variations in the $h$ and $k$ spacing become increasingly correlated, and the magnitudes of the variances increase rapidly. In other words, the broadening observed contains components along both the crystallographic \textbf{a} and \textbf{b} axes. 

Figure 5b illustrates that the Stephens parameters corresponding to the only the \textbf{b}-axis (S$_{040}$), and those with no contribution at all from the \textbf{b}-axis are low in magnitude, and show no temperature dependencies. Figure 5c illustrates that while there is a correlation between the monoclinic \textbf{a}- and \textbf{b}-axes in the behaviour of the S$_{220M}$ parameter, it is dwarfed by the behaviour of the parameters in which contributions from the \textbf{b}- and \textbf{c}-axes are present: S$_{121M}$ and S$_{022M}$. 

Figures 5a-c thus reveal that the diffraction data contains contributions to the broadening of the peaks which is anisotropic in nature, and that the most significant contributions are those of the S$_{022}$ and S$_{121}$ parameters, while the tetragonal structure shows no significant correlation in the equivalent parameters. This data is therefore in line with the data of Figure 4, and Tables 1 and 2, and supports the hypothesis that the energy barrier between the structures corresponds to the compression and expansion of the characteristic distance of Figure 3d.

The observed behaviour of the S$_{220}$ data is in some respects not surprising, as when plotted as a function of temperature, it is basically an un-normalized temperature correlation function of the variances of the \textbf{a*} and \textbf{b*} axes. The divergence of this correlation at the T$_{c}$ point is in line with critical behaviour expected at a phase transition, however, we do not attempt to explore this aspect in this work.

\begin{figure*}[]
  \includegraphics[width=1.9\columnwidth]{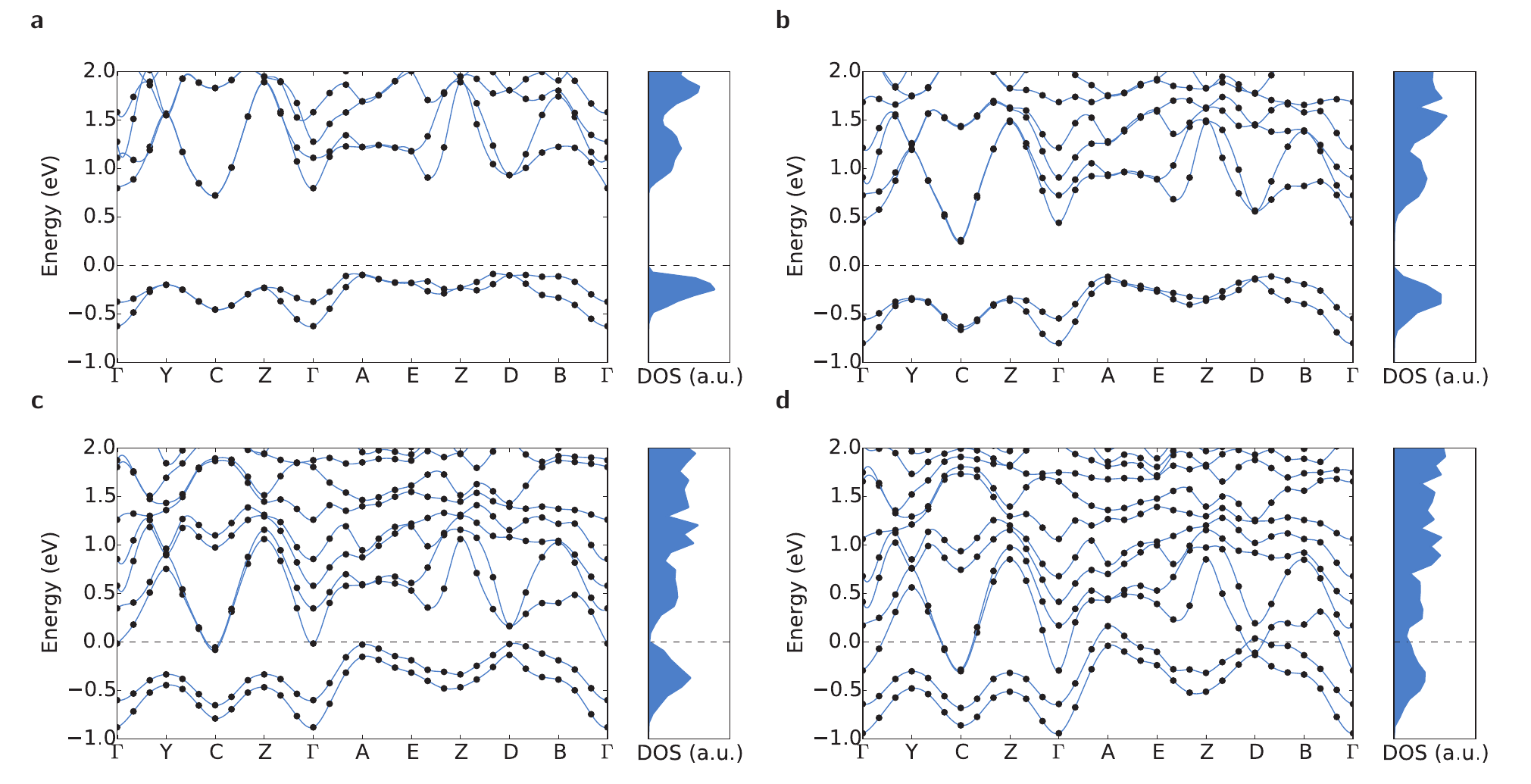}
  \caption{GW band eigenvalues (filled black circles fitted with blue splines to guide the eye, left panel) and electronic densities of states (filled blue curve, right panel) of a) the M$_{1}$ structure, b) the ``Step 1" structure, c) the ``Step 2" structure and d) the ``Step 3" structure. The Step numbers correspond to the total energy points in Figure 2.}
\end{figure*}
\subsection*{Electronic Structure}
What remains to be determined however, is the effect of these structural rearrangements on the electronic structure. There is little question that the electronic structures of the tetragonal and monoclinic forms are metallic and insulating respectively,\cite{Zheng2015} however of interest in this study is the behaviour of the band structure across this structural phase transition, in order to determine whether the structural and electronic phase transitions are intrinsically related, or in fact merely coincident. The next section explores this in detail.

The band structures of the monoclinic ground state, and structures ``Step 1", ``Step 2" and ``Step 3" calculated using the GW approximation (black filled circles, fitted with blue splines) are presented in Figure 6. The densities of states are plotted next to each band structure, on the same energy scale (blue filled curve). As expected, the monoclinic structure is insulating, with a band gap of $\sim$ 0.70 eV, in excellent agreement with experiments (0.70 eV).\cite{Shin1990} 

However as the structure transitions to the slightly more symmetric forms of Step 1 and Step 2 the densities of states indicate that the gap closes, and the structure becomes metallic. The corresponding band structures indicate that this occurs \textit{via} two simultaneous mechanisms. The dispersions of the valence bands in the $\Gamma$ $\rightarrow$ \textit{A} direction suggests that in comparison to the band minima at $\Gamma$, the higher energy states near E$_{F}$ are shifting upwards, most significantly near \textit{A} and \textit{D}. At the same time, the conduction band minima at $\Gamma$ shift downwards. In structure ``Step 2" the conduction and valence bands overlap (this was determined by inspection of the charge density), and the indirect gap closes. The band structure of Figure 6c, when compared with the total energy data of Figure 2, indicates that the electronic band gap destabilises before the top of the energy barrier is crested.

We can gain a better idea of how the structural transitions are affecting these states by transforming them to real space charge densities and comparing them. Figures 7a-b present charge density isosurfaces of the valence and conduction band states at $\Gamma$ in the (0$\bar{1}$1) plane of the ground state monoclinic structure respectively, while Figures 7c-d present charge density isosurfaces in the (0$\bar{1}$1) plane of the valence and lowest energy conduction band states at the \textit{D} point of the M$_{1}$ structure. 

From comparison of Figures 7a-b, it is obvious that the valence band state at $\Gamma$ consists of shared charge density between the vanadium (grey spheres) atoms, while the conduction band state corresponds to isolated density on each vanadium atom. This suggests that the gap between the valence and conduction bands arises from bonding/antibonding splitting. The same story is repeated at $D$. The valence band state contains density linking the Peierls paired vanadium atoms, while the conduction band state consists again of isolated density on each vanadium atom.  The band structure data of Figure 6 indicate that as the structure transitions away from the M$_{1}$ form and towards the tetragonal, the splitting between these states decreases considerably, with the eigenvalues at $D$ crossing over by Step 3. 

This charge density, combined with the eigenvalues, and the inter-vanadium spacing data of Figure 3 indicate that as the structural phase transition progresses, starting from the M$_{1}$ structure, the inter-vanadium distance of the Peierls pairs increases, destabilising bonding states with respect to anti-bonding states, narrowing the gap between the conduction and valence bands, until it disappears completely and the structure becomes metallic. 

Figure 8 plots the value of the local potential along a line segment connecting the Peierls paired vanadium atoms of the M$_{1}$, ``Step 3", ``Step 6" and tetragonal structures, and from the data, it is obvious that as the inter-vanadium distance increases, the height of the potential barrier between the nuclei also increases. This increase will significantly affect the wavefunctions of the highest energy electrons which are obviously less tightly bound, resulting in less electron sharing between the paired atoms, consequently raising the energy of bonding configurations with respect to anti-bonding.

\begin{figure}[th!]
  \includegraphics[width=\columnwidth]{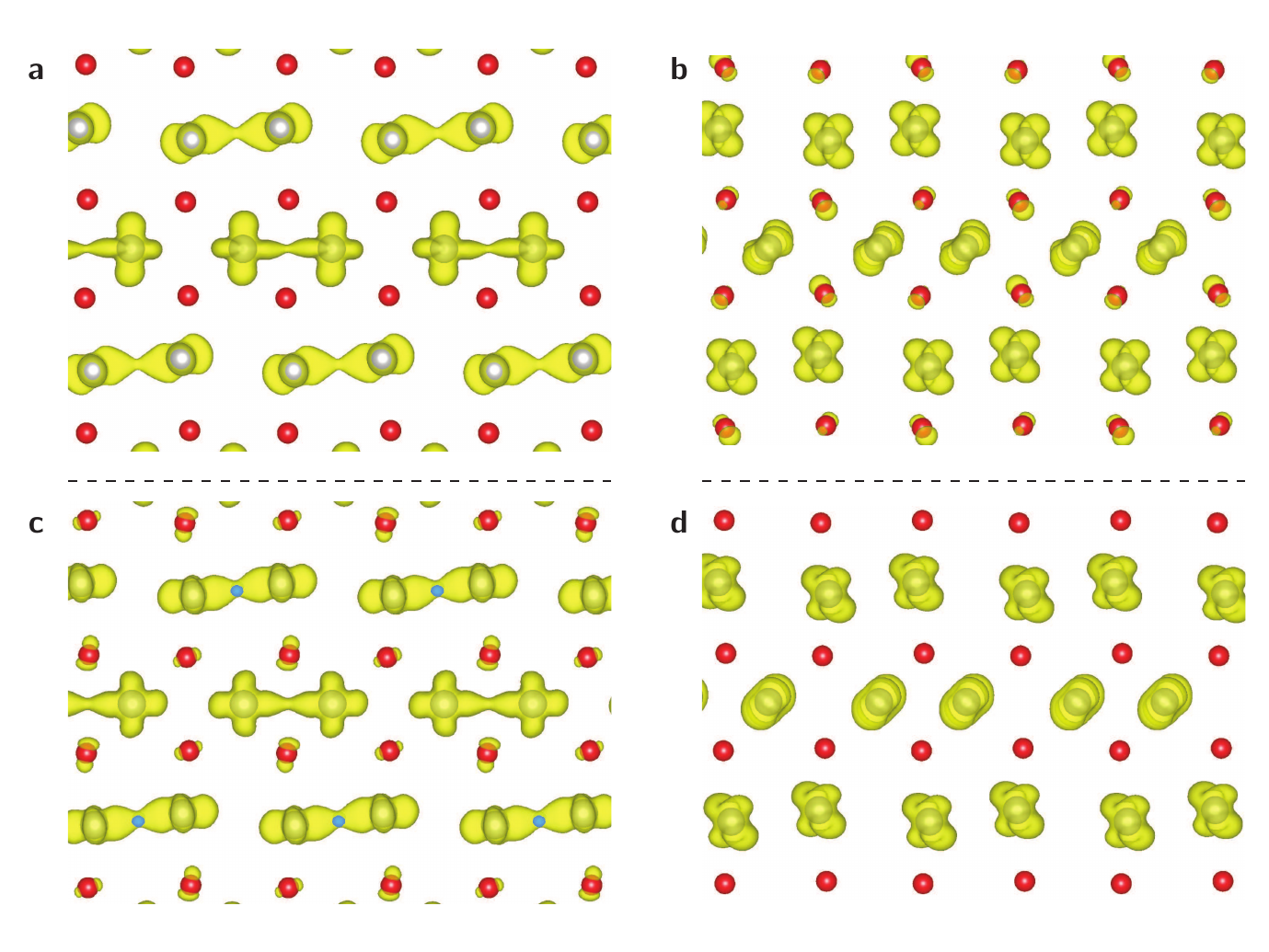}
  \caption{Charge densities in the (0$\bar{1}$1) plane of a) the highest valence band state at $\Gamma$, b) the lowest conduction band state at $\Gamma$, c) the highest valence band state at $D$ and d) the lowest conduction band state at $D$. The conduction and valence band states correspond to bonding and anti-bonding configurations respectively, and thus as the gap narrows between them across the transition (see Figure 6) this indicates the bonding-anti bonding splitting is being reduced which results in the closing of the electronic band gap. Vanadium atoms are gray, while oxygen atoms are red.}
\end{figure}

\begin{figure}[h!]
  \includegraphics[width=0.8\columnwidth]{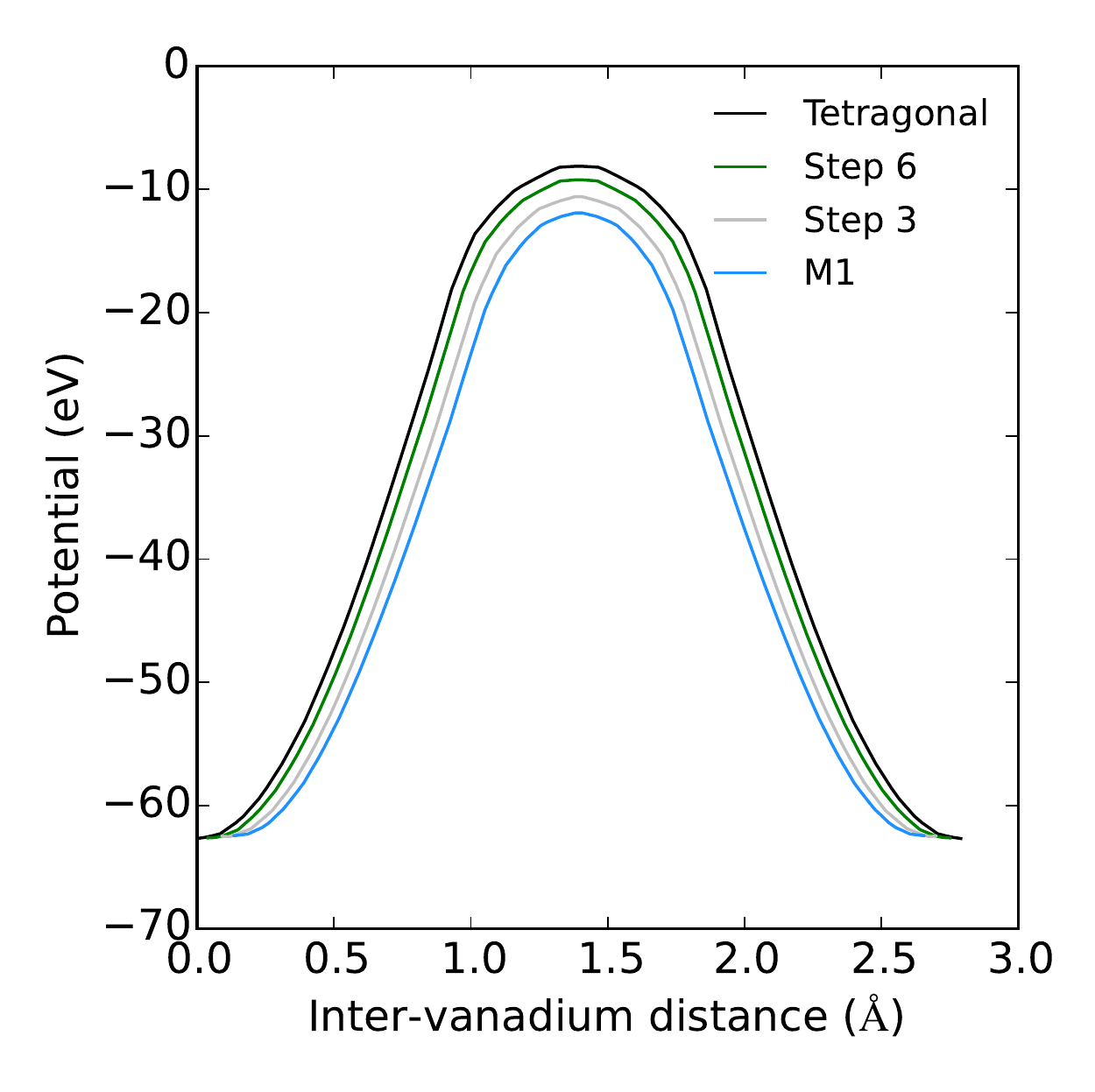}
  \caption{Total local potentials (eV) along a line segment connecting the Peierls paired vanadium atoms of the M$_{1}$, Step 3, Step 6 and tetragonal structures. The inter-vanadium distances have been shifted such that the midpoints of each segment coincide, in order to align the potential barriers.}     
\end{figure}

\section*{Discussion}

A more complete picture of the processes occurring across the VO$_{2}$ structural/electronic phase transition can now be pieced together using the data presented. The barrier which separates the structures is a consequence of the need to compress the inter-vanadium spacing along the [011] and [01$\bar{1}$] directions of the monoclinic structure in order to effect the structural phase transition. This mirrors the appearance of the M$_{2}$ structure of VO$_{2}$ upon the input of strain along the [110]$_{T}$ direction.\cite{Pouget1975} The chains consist of a linear, Pe  These displacements result in the destabilisation of the bond/anti-bonding splitting of the monoclinic structure, due to the increase in the potential barrier between the Peierls paired vanadium atoms. This results in the conduction and valence bands overlapping; an insulator-metal transition.

If the energy barrier is indeed a consequence of the atomic motion of Figure 3d, then attempts to manipulate the transition temperature which involve modulation of the stress or strain along the [110]$_{T}$ and [0$\bar{1}$1]$_{T}$ directions would produce the most significant effects. This is consistent with X-ray absorption studies\cite{Booth2009a} which indicate that the depression of the transition temperature by tungsten-doping correlates with an increase in the V-V corner spacing. If the energy increase of Figure 2 is a consequence of electronic repulsion, then increasing the V-V corner distances will lower the repulsion due to contraction as the increased internuclear separation will result in a lowering of the electrostatic potential between the atoms, reducing the barrier height.

\section*{Methods}

Variable temperature Synchrotron X-ray Powder Diffraction was conducted at the Australian Synchrotron Powder Diffraction Beamline. Samples were sealed in 0.3 mm borosilicate glass capillaries. Prior to data collection, the wavelength was set at ~0.82732 $\text{\AA}$ using a Si (111) double crystal monochromator.  The exact wavelength was refined using the NIST 660b LaB$_{6}$ standard reference material. A Cyberstar hot-air blower was used to control the temperature to within 0.1 $-$ 0.2 $^\circ$C at each data collection temperature. Traces were recorded for 5 minutes at each of the two detector settings after the sample had reached the set point temperature and equilibrated for 10 minutes.

Quantitative Rietveld analysis was performed on the data using the Bruker TOPAS\textsuperscript{TM} V4.2 program to determine the weight percentage of phases present.  Background signal was described using a Chebyshev polynomial linear interpolation function.  A broad pseudo-Voight function was also used to model the background contribution form the capillary. Cell parameters, atom positions, (tightly constrained) isotropic thermal parameters, Gaussian and Lorentzian contributions to peak full widths at half maximum and scale factor were all refined.

The anisotropic broadening of the diffraction peaks observed in Figure 4 was hypothesized to originate from one of two sources. Either there was some anisotropy in the crystallite shapes, leading to broadening of lines corresponding to directions in which fewer planes are stacked, or the crystal grains exhibit a distribution of residual strains originating from the transition from tetragonal to monoclinic. Thermal effects were deemed an unlikely origin, as refinements produced similar thermal parameters at all temperatures, and the anisotropic broadening observed is \textit{smaller} at higher temperature. In addition, the Debye-Waller factor\cite{Debye1913,Waller1923} tends to reduce the scattered intensity, however in comparisons between the equivalent monoclinic and tetragonal peaks, such as (011)$_{M}$ and (110)$_{T}$ the peaks integrate to the same total intensity. Employing Jarvinen's method\cite{Jarvinen1993} to account for anisotropic broadening led to rather poor fits in comparison to those generated by the Stephens method\cite{Stephens1999} for strain broadening, indicating that while some crystallite size ansiotropy may exist, the broadening is dominated by the strain distribution.

Strain analysis of the X-ray data was performed using the method developed by Stephens,\cite{Stephens1999} which is a phenomenological approach to determining the contributions to broadening induced by anisotropic variations in plane spacings. We repeat the central thesis of this approach here, but for a more complete treatment the reader is referred to the original work.\cite{Stephens1999}

The spacing of planes with Miller indices \textit{hkl} is given by:
\begin{equation}
\frac{1}{d^{2}}=M_{hkl}=Ah^{2}+Bk^{2}+Cl^{2}+Dkl+Ehl+Fhk
\end{equation}
Re-labeling the metric parameters $\{A,...,F\}$ as $\{\alpha_{i}\}$ and assuming that they have Gaussian distributions characterised by a covariance matrix $C_{i,j}=\langle(\alpha_{i}-\langle\alpha_{i}\rangle)(\alpha_{j}-\langle\alpha_{j}\rangle)\rangle$, the variance of $M_{hkl}$ can be written:
\begin{equation}
\sigma^{2}(M_{hkl})=\sum_{i,j}^{}C_{ij}\frac{\partial M}{\partial \alpha_{i}}\frac{\partial M}{\partial \alpha_{j}}
\end{equation}
which since $\partial M/\partial \alpha_{1} = h^{2}$, $\partial M/\partial \alpha_{5} = hl$ \textit{etc.} can be re-written:
\begin{equation}
\sigma^{2}(M_{hkl})=\sum_{HKL}^{}S_{HKL}h^{H}k^{K}l^{L}
\end{equation}
where from equations (1) and (2) the terms S$_{HKL}$ are obviously defined for $H+K+L=4$. The contribution from anisotropic strain broadening to the full-width-half-maximum (FWHM) of a diffraction line can be written using the Bragg equation and (4) as:
\begin{equation}
\Gamma_{A}=[\sigma^{2}(M_{hkl})]^{1/2}\frac{tan\theta}{M_{hkl}}
\end{equation}
This $\Gamma_{A}$ is combined with the usual parameters for Gaussian and Lorentizian line-widths to give expressions for anisotropically broadened line-shapes which are fitted to the experimental data, and the S$_{hkl}$ are extracted from the fit. The Gaussian and Lorentzian broadening parameters of Tables I and II were extracted from fits to the individual peaks presented in the data of Figure 4.

\subsection*{Force calculations}
The monoclinic\cite{Andersson1954} and tetragonal\cite{Marezio1971} structural parameters were input to DFT geometry relaxations using the VASP code\cite{Kresse1996} and the Generalized Gradient Approximation to exchange and correlation of Perdew et al.,\cite{Perdew1996} on 6$\times$6$\times$6 and 8$\times$8$\times$6 Monkhorst-Pack\cite{Monkhorst1976} k-space grids. The structures were then relaxed to their respective ground states using Methfessel and Paxton smearing\cite{Methfessel1989} and the conjugate gradient algorithm. Upon reaching the desired ground states, a 1$\times$1$\times$2 supercell of the tetragonal structure was constructed in order to have the same dimensions as the monoclinic form. 

The Cartesian atomic positions of the tetragonal structure were then subtracted from those of the monoclinic structure, which generated vectors describing the movement of the atoms across the transition. Vectors describing the changes in unit cell dimensions were obtained in the same manner.  These vectors were then divided such that 10 structures were generated, with the monoclinic structure being the first, and the tetragonal being the last and the intermediate structures are labelled ``Step 1" to ``Step 8". The elastic band technique\cite{Henkelman2000} was then applied to these structures, in order to find the minimum energy path between them. The use of DFT to determine the total energies of Figure 2 from the structures optimised by the elastic band method, rather than DFT+U or hybrid functionals stems from the requirement to maintain a consistent Hamiltonian for the calculation of the energies along the minimum energy path as the energy landscape, by definition, will be Hamiltonian dependent. 

\subsection*{Electronic Structure}
The relaxed structural parameters were used as input to Density Functional Theory\cite{Kohn1965,Kresse1996} calculations on $6\times 6\times 6$ Monkhorst-Pack k-space grids, again using the Generalized Gradient Approximation (GGA) approach to exchange and correlation of Perdew \textit{et al.}\cite{Perdew1996} with the Brillouin zone integration approach of Bloechl \textit{et al.}\cite{Bloechl1994} Frequency-dependent GW calculations\cite{Hedin1965} were performed using the implementation of Shishkin and Kresse\cite{Shishkin2006} in VASP.\cite{Kresse1996} The GW calculations were performed using a grid of 50 frequency points and an energy cutoff of 200 eV.

\section*{Acknowledgements}

This work was supported by computational resources provided by the Australian Government through the National Computational Infrastructure under the National Computational Merit Allocation Scheme. DWD acknowledges the support of the ARC Centre of Excellence for Nanoscale BioPhotonics (CE140100003). JMB, AJS and PSC thank the Australian Synchrotron Research Program for continued support. 

\end{document}